\documentclass[letterpaper]{sfchem}
\usepackage{graphicx}

\begin{document}

\title{Molecular abundances in low-mass protostellar envelopes}

\author{Jes K. J{\o}rgensen\inst{1} \and Fredrik L. Sch\"{o}ier\inst{1} \and Ewine F. van Dishoeck\inst{1}} 
\institute{Leiden Observatory, P.O.\ Box 9513, 2300 RA Leiden, The Netherlands} 
\authorrunning{J{\o}rgensen, Sch\"{o}ier and van Dishoeck}
\titlerunning{Molecular abundances in low-mass protostellar envelopes}

\maketitle 

\begin{abstract}
A study of the chemical structure of the envelopes around a sample
protostars is introduced. Physical models for the envelopes derived
using 1D radiative transfer modeling of their dust continuum emission
are used as input for Monte Carlo modeling of single-dish line
observations in order to establish the chemical inventories for the
different sources. CO and HCO$^+$ abundances are found to be
correlated with envelope mass: the sources with the most massive
envelopes show the lowest abundances, supporting the idea that the
depletion of these molecules is most efficient in the colder, denser
environments. Other molecules like CS and HCN do not show a similar
trend. Deuterated species like DCO$^+$ and DCN also show signs of
correlations with envelope mass and temperature: the deuterium
fractionation of HCO$^+$ is most prominent for the coldest and most
massive envelopes in the sample and at the same time anti-correlated
with the fractionation of HCN. This puts constraints on the low
temperature gas-phase deuterium chemistry.


\end{abstract}

\section{Introduction}
Understanding the chemistry of protostellar environments is of great
importance for addressing topics of star formation. It is directly
related to processes regulating star formation, for example through
the ionisation and the chemistry also potentially serves as an
evolutionary clock against which physical models, e.g. for the
protostellar collapse, can be compared (see e.g. \cite*{vandishoeck98}
for an overview). Recent evidence suggests that many of the
traditional molecular tracers of density and temperature may be
subject to significant gradients in abundances. Examples are the
freeze-out of molecules like CO onto dust grains both in pre-stellar
cores (\cite{caselli99}, \cite{bacmann02}) and in the cold and dense
parts of low-mass protostellar envelopes (\cite{jorgensen02}) and
enhancement of molecules due to liberation of ice mantles in regions
of higher temperatures similar to ``hot cores'' or in regions subject
to shocks driven by protostellar outflows (\cite{bachiller97},
\cite{ceccarelli00a}, \cite{schoeier02}).

In this paper a preliminary overview of the results from a large
survey of the chemical properties of a sample of low-mass protostars
is presented. For further details see \cite*{jorgensen02} and
J{\o}rgensen et al. (2003; in prep.).

\section{Sample, observations and modeling}
The sample of objects in this study comprises 18 pre- and protostellar
objects, including 11 class 0 objects, 5 class I objects and 2
pre-stellar cores; the full sample is presented in
\cite*{jorgensen02}. These objects have been observed in a range of
molecular transitions using the 15~m James Clerk Maxwell Telescope
(JCMT), 20~m telescope at Onsala Space Observatory and the IRAM Pico
Valeta 30~m telescope between 2000 and 2002.

The physical structure of each envelope is derived using 1D radiative
transfer calculations. Power-law density profiles are assumed for a
range of envelope parameters (e.g. density slope and optical
thickness) and the corresponding temperature profiles calculated using
the radiative transfer code, DUSTY (\cite{dusty}) for a given central
source of heating. For each model, images are constructed and compared
to the observed SCUBA images (450 and 850~$\mu$m) and SEDs over the
range from 60~$\mu$m to 1.3~mm. It is found that each protostellar
envelope can successfully be modeled under these assumptions, whereas
the pre-stellar cores can not, which is not unexpected since these
sources per definition are characterized as not having central sources
of heating.

\begin{figure}[ht]
\resizebox{\hsize}{!}{\includegraphics{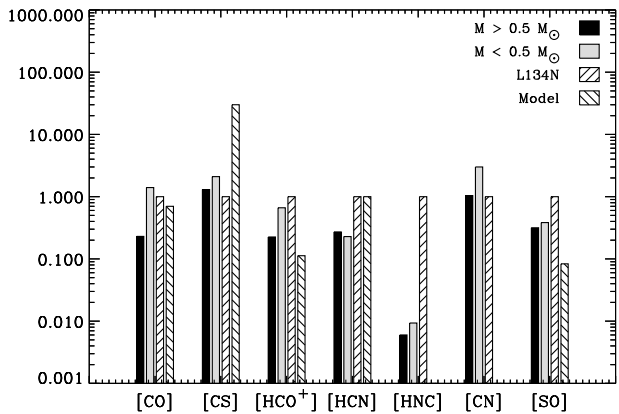}}
\caption{Comparisons between average abundances with respect to H$_2$
for different molecules for objects with envelope masses larger or
smaller than 0.5~$M_\odot$, and with the abundances towards the ``C''
position in the dark cloud L134N (Dickens et al. 2000) and the
abundances in the stage before the collapse in the chemical models of
Bergin \& Langer (1997). All abundances normalized to the abundances
of L134N. (From J{\o}rgensen et al. 2003; in prep.)}\label{abundhist}
\end{figure}

The derived physical models are subsequently used as basis for
calculating the molecular excitation and radiative transfer for the
molecular lines using the Monte Carlo code of
\cite*{hogerheijde00vandertak}. The fractional abundances (assumed to
be constant) are constrained through comparisons with the integrated
line intensities and it is found that the majority of the lines can
successfully be accounted for with the presented models and
assumptions. In Fig.~\ref{abundhist}, an overview of the average
abundances for the class 0 and I objects are compared to the molecular
cloud L134N (\cite{dickens00}) and the abundances in the steady state
or pre-collapse phase of the models of \cite*{bergin97}. The
abundances are, where possible, derived from the minor isotopic
lines. The main isotopic lines of e.g. CS and HCO$^+$ are more
sensitive to departures from spherical symmetry or the interaction of
molecular outflows with the envelope material.

\section{Molecular depletion}
\begin{figure}[ht]
\centering
\resizebox{7.5cm}{!}{\includegraphics{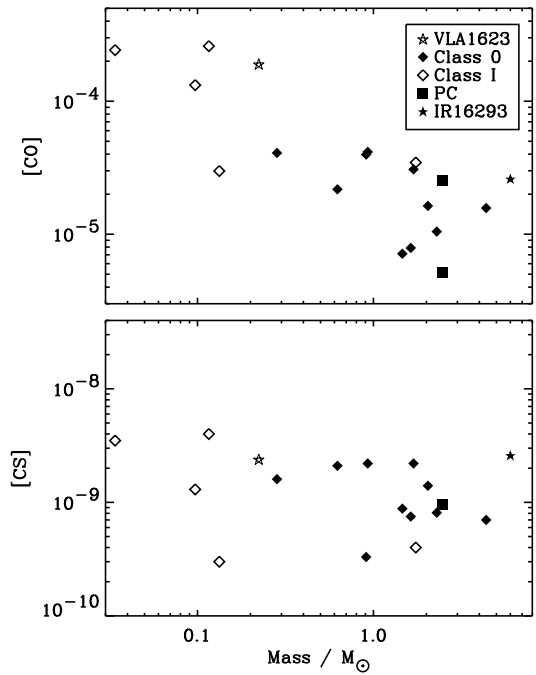}}
\caption{CO (upper panel) and CS (lower panel) abundances as functions
of mass. (From J{\o}rgensen et al. 2002, 2003; in prep.)}\label{co}
\end{figure}

\begin{figure}[ht]
\centering
\resizebox{7.5cm}{!}{\includegraphics{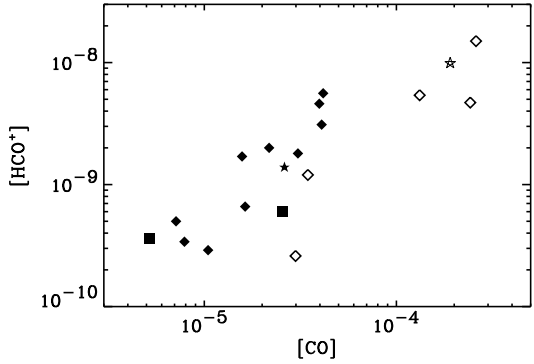}}
\caption{HCO$^+$ abundance as a function of CO abundance. Symbols as
in Fig.~\ref{co}. (J{\o}rgensen et al. 2003; in prep.)}\label{hcop}
\end{figure}

As can be seen from Fig.~\ref{abundhist} and \ref{co} significant
differences exist between the various molecules relative to the two
comparison sets of abundances and also between the samples of class 0
and I objects for some molecules -- in particular CO and HCO$^+$. CO is
thought to be frozen out onto dust grains at low temperatures ($T \leq
30$~K ; \cite{jorgensen02}) and CO is indeed most depleted for the
sources with the most massive envelopes (Fig.~\ref{co}), i.e., the
envelopes with a region in which the temperatures are low enough and
densities high enough that the depletion can be effective. The primary
formation route of HCO$^+$ is connected to the presence of CO through
the reaction ${\rm CO}+{\rm H}_3^+ \rightarrow {\rm HCO}^+ + {\rm
H}_2$, as is indeed illustrated by the data in Fig.~\ref{hcop}.

Other molecules like CS and HCN do not show a similar trend with mass,
which at first may seem unexpected (lower panel of Fig.~\ref{co}): CS
depletion is expected to occur at even lower densities than for CO and
CS should stay bound to the dust grains to even higher temperatures
than CO due to its larger binding energy to the dust
grains. Comparison to the models of \cite*{bergin97} shows that the CS
abundances observed in the protostellar environments are indeed lower
by more than an order of magnitude, which could indicate that the
observed lines only probe the region where CS is depleted. A detailed
comparison with other sulfur-bearing species like SO will be needed to
fully address this, and similarly for the nitrogen-bearing species
that also show interesting features -- in particular the strikingly low
HNC abundances for the protostars compared to other star-forming
environments.

\section{Deuterium fractionation}
\begin{figure}[ht]
\centering
\resizebox{7.5cm}{!}{\includegraphics{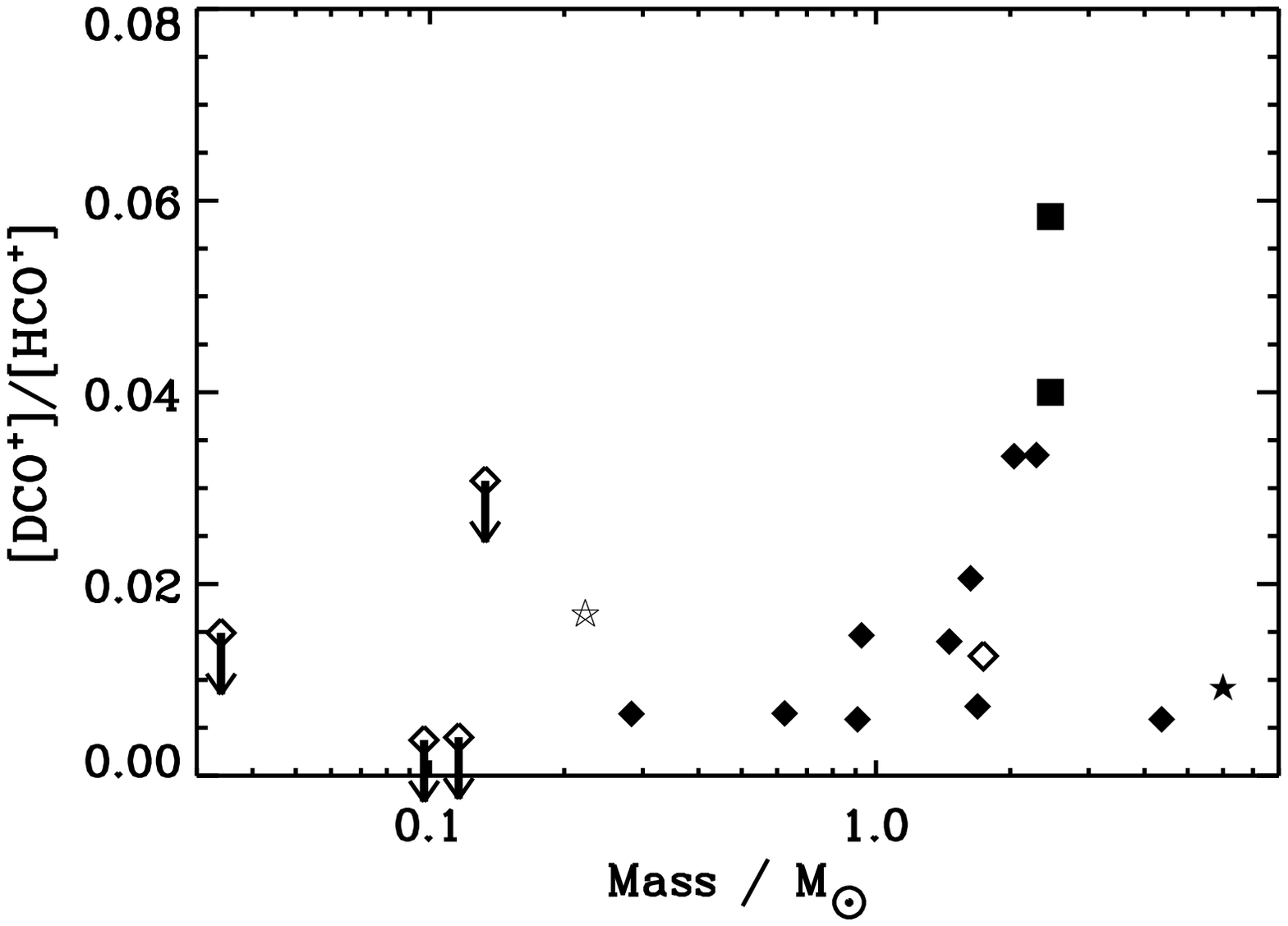}}
\resizebox{7.5cm}{!}{\includegraphics{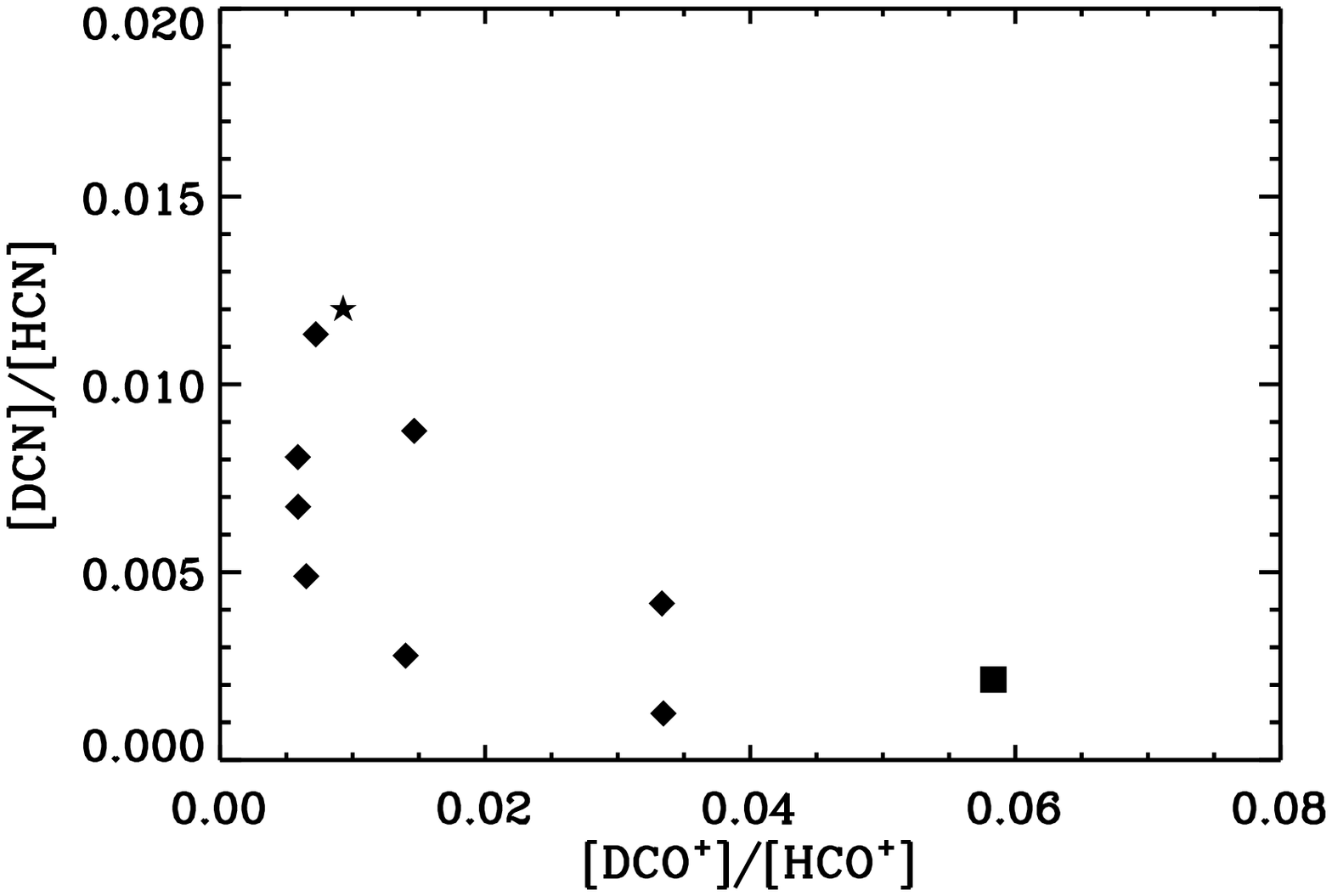}}
\caption{[DCO$^+$]/[HCO$^+$] ratio plotted against mass (upper panel)
and [DCN]/[HCN] plotted against [DCO$^+$]/[HCO$^+$] ratios (lower
panel). Symbols as in Fig.~\ref{co}. (J{\o}rgensen et al. 2003; in
prep.)}\label{dcop}
\end{figure}

Another interesting aspect of the chemistry in the envelopes around
protostars is the degree of deuterium fractionation. Gas-phase
deuterium fractionation of molecules like HCO$^+$ is expected to be
particularly efficient at low temperatures (e.g. \cite{roberts00a})
due to small zero-point energy differences of reactions like ${\rm HD}
+ {\rm H}_3^+ \rightarrow {\rm H}_2{\rm D}^+ +{\rm H}_2$. As
illustrated in Fig.~\ref{dcop}, the highest degree of deuterium
fractionation is seen to occur in the pre-stellar cores and class 0
objects with low temperatures ($T \leq 20$~K) and high
densities. Interestingly, the HCN deuterium fractionation seems to be
anti-correlated with that of HCO$^+$ as shown in the lower panel of
Fig.~\ref{dcop}: for HCN deuterium fractionation through other species
like ${\rm CH}_3^+$ (i.e. forming ${\rm CH}_2{\rm D}^+$) could become
increasingly important in producing DCN at temperatures $T \sim 30$~K
(e.g. \cite{turner01}). Such a scenario could explain the observed
anti-correlation between the deuteration of HCO$^+$ and HCN.

%
\begin{acknowledgements}
The authors are grateful to the Leids Kerkhoven-Bosscha Fond for
supporting their participation in the conference financially.
\end{acknowledgements}

\end{document}